\documentclass[sigconf,10pt,nonacm]{acmart}
\usepackage{custom}

\title{Faster Offloads by Unloading them -- The RDMA Case}

\author{Georgia Fragkouli}
\affiliation{%
  \institution{ETH Z\"urich}%
  \city{Z\"urich}
  \country{Switzerland}%
}
\author{Laurent Vanbever}
\affiliation{%
  \institution{ETH Z\"urich}%
  \city{Z\"urich}
  \country{Switzerland}%
}  

\begin{abstract}

From hardware offloads like RDMA to software ones like eBPF, offloads are everywhere and their value is in performance.
However, there is evidence that fully offloading---even when feasible---does not always give the expected speedups.
Starting from the observation that this is due to changes the offloads make---by moving tasks from the application/CPU closer to the network/link layer---we argue that to further accelerate offloads, we need to make offloads reversible by \emph{unloading} them---moving back part of the offloaded tasks.

Unloading comes with a set of challenges that we start answering in this paper by focusing on (offloaded) RDMA writes:
which part of the write operation does it make sense to unload? how do we dynamically decide which writes to execute on the unload or offload path to improve performance? how do we maintain compatibility between the two paths?
Our current prototype shows the potential of unloading by accelerating RDMA writes by up to 31\%.
 
\end{abstract}

\begin{document}

\maketitle

\section{Introduction}
\label{sec:intro}

\emph{``If you can offload, do it''} seems to be common knowledge nowadays, whether it is to offload simple computations, like checksumming, to NICs; packet filtering logic to the kernel with eBPF; or entire network stacks to RDMA NICs (RNICs).
No matter what is offloaded, the goal is always the same: increasing performance by improving throughput and/or the latency of operations. 

Yet, while offloading does generally increase performance, it can also decrease it. One such example is offloading TCP/IP processing to the NIC, which can end up exhausting the NIC memory and lead to
worse performance for surplus connections~\cite{mogul04unveilingtransport,kim06tcpoffloadhandoff}, and which led the Linux Foundation to opt out of supporting it~\cite{linuxfoundationtoe}.

This triggers the question: \textit{when are offloads beneficial, and what causes them to underperform?}
Intuitively, the performance of an offload depends on the framework/hardware that runs it. Specifically, offloading a task changes its execution by adapting: \emph{where} the task is executed (e.g., the NIC instead of the CPU), \emph{how} the task is executed (e.g., with kernelspace instead of userspace optimization strategies), and \emph{what} resources are available (e.g., NIC cache instead of DRAM). This adds performance overheads that for certain workloads negate the benefits of (full) offloads.

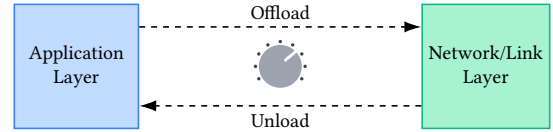
\begin{figure}
    \centering
\resizebox{0.85\linewidth}{!}{%
\begin{tikzpicture}[
  module/.style={rectangle, draw, thick, minimum width=2.2cm, minimum height=2.2cm},
  submodule/.style={rectangle, draw, thick, minimum width=10cm, minimum height=1.2cm, fill=blue!5},
  decsubmodule/.style={rectangle, draw, thick, minimum width=1.2cm, minimum height=2.2cm, fill=blue!5},
  arrow/.style={thick, -{Latex[length=3mm, width=2mm]}}
]
  \definecolor{blue-300}{RGB}{147,197,253}
  \definecolor{emerald-300}{RGB}{110,231,183}
  \definecolor{amber-300}{RGB}{253,224,71}
  \definecolor{red-300}{RGB}{252,165,165}
  \definecolor{purple-300}{RGB}{216,180,254}
  \definecolor{pink-300}{RGB}{251,207,232}
  \definecolor{cyan-300}{RGB}{103,232,249}
  \definecolor{yellow-300}{RGB}{253,230,138}
  \definecolor{red500}{RGB}{239,68,68}
  \definecolor{green-500}{RGB}{34,197,94}
  \definecolor{green-300}{RGB}{134,239,172}

  \definecolor{blue-200}{RGB}{191,219,254} 
  \definecolor{emerald-200}{RGB}{167,243,208} 
  \definecolor{blue-500}{RGB}{59,130,246} 
  \definecolor{emerald-500}{RGB}{16,185,129} 
  \node[module, fill=blue-200, draw=blue-500, thick, align=center] (app) {Application\\Layer};
  \node[module, fill=emerald-200, draw=emerald-500, thick, align=center, right=5cm of app] (net) {Network/Link\\Layer};

  \path (app.east) ++(0,0.7) coordinate (appUp) (net.west) ++(0,0.7) coordinate (netUp);
  \draw[arrow, dashed] (appUp) -- node[midway, above] {Offload} (netUp);
  \path (net.west) ++(0,-0.7) coordinate (netDown) (app.east) ++(0,-0.7) coordinate (appDown);
  \draw[arrow, dashed] (netDown) -- node[midway, below] {Unload} (appDown);

  \coordinate (mid) at ($(app.east)!0.5!(net.west)$);
  \definecolor{gray-400}{RGB}{156,163,175} 
  \def\knobradius{0.38}
  \fill[gray-400] (mid) circle [radius=\knobradius];
  \definecolor{gray-700}{RGB}{55,65,81} 
  \foreach \a in {330,0,30,60,90,120,150,180,210} {
    \fill[gray-700] (mid) ++(\a:{\knobradius+0.06}) circle [radius=0.03];
  }
  \draw[line width=1.5pt, white, line cap=round] ($(mid)+(0.09,0.11)$) -- ++(40:0.28);
\end{tikzpicture}
}%
    \caption{\sys makes offload operations reversible and dynamically decides whether to unload or offload.}
    \label{fig:bidirectional}
\end{figure}

Given that offloading is not always beneficial, we argue that one should be able to \emph{unload} the offloads, i.e., revert the aforementioned changes to avoid performance degradation when the
workload is unfavorable for the offload.

Unloading comes with its own challenges.
First, since offloading \emph{is} performant for many workloads, one has to carefully and \emph{dynamically} decide which tasks to unload to optimize performance.
Second, one has to carefully decide \emph{how} to unload, as naively fully unloading a task would often do more harm than good.
For example, from a performance point of view, it would never make sense to unload RDMA operations by reverting them to the much slower TCP stack~\cite{zhu15congestioncontrolrdma}.
Third, the need to dynamically unload means that rewriting applications to use unloading is not only an overhead but an ineffective one in improving performance.
So, if the application code remains the same but we dynamically replace offloaded application calls with unloaded ones, how do we maintain compatibility with application expectations?

We start solving these challenges by focusing on the problem of efficient remote data transfers.
RDMA is the de-facto technology in the cloud for improving the performance of remote transfers from/to disaggregated storage. (In Azure, 70\% of the traffic is RDMA~\cite{bai23azurerdma}.)
RDMA achieves that by changing the \emph{where}---it moves the handling of memory transfers from CPUs and the TCP/IP stack to RNICs---and the \emph{what}---it relies on RNIC caches instead of CPU resources.
These changes mean that when RNICs need to access host memory, they must go over the peripheral interconnect (e.g., PCIe).
The result is a well-known overhead under cache misses for virtual-to-physical memory address translations~\cite{guo16rdmaatscale,kong23rdmaisolation}: when an RDMA write operation arrives at the remote RNIC and the respective translation is not in the RNIC's TLB, called Memory Translation Table (MTT) cache, the write must wait for the translation to be fetched over the PCIe---inflating latency by 2x (\cref{sec:motivation}) and degrading throughput by 40\%~\cite{kong23rdmaisolation}.

Our approach to improving RDMA write latency, \sys, is to make offloading \emph{bidirectional} (\cref{fig:bidirectional}): allow to dynamically switch between (offloaded) RDMA writes and their unloaded counterparts.
Besides the original offload path, \sys has an unload path that moves the task of writing to the target memory address from the RNIC back to the CPU.
This makes the unload path more performant as it frees up the RNIC from RDMA'ing to memory addresses whose translations are not in the MTT cache, avoiding slow translation resolutions over the PCIe.
Specifically: the unload path redirects RDMA writes to special, memory regions (e.g., in a pool of large pages), whose translations are in the MTT cache.
The writes are buffered in these regions temporarily until the CPU takes over copying the data to the target memory addresses.
Equipped with the two paths, \sys accelerates RDMA writes by adaptively switching between paths.
The idea is to keep on the offload path only writes whose characteristics allow them to benefit from the RNIC cache.

We evaluate \sys by building a preliminary prototype using NVIDIA ConnectX-5 Ex RNICs.
It showcases that \sys improves 
latency by up to 31\%
(\cref{fig:effectmrszipfian} and \cref{sec:eval}).
These gains match our intuition: 1)~the CPU sits closer to host memory, thus resolves translations faster than the RNIC, and 2)~the small overhead of copying small buffers is over-compensated by cutting on PCIe traversals.

\section{Motivation}
\label{sec:motivation}

To gain performance, offloads impact execution and performance by changing: 1)~\emph{where} operations are executed, 2)~\emph{how} they are executed, and 3)~\emph{what} resources they use.
These changes impose costs and tradeoffs.
Offloads are designed so that these tradeoffs typically lean heavily on the gain side.
The question is: how do we ensure that offloads always get every last bit of performance?
This is challenging as:

\paragraph{Problem 1: Offload performance is sensitive to workload characteristics}
The same reasons that make offloads performant in the common case add penalties under certain workloads:
First, for offloads that change the execution location, the assumption is that any extra traversals between the offload location and resources close to the non-offload execution location are infrequent; otherwise, the location change adds overheads.
For example, for RDMA writes, instead of the CPU (loc 1), it is the RNIC (loc 2) that initiates virtual-to-physical address translations.
When fetching translations becomes significant, the location change degrades latency.

Second, for offloads that change the execution, the assumption is that the offloaded application benefits little from optimization strategies left behind.
For example, eBPF~\cite{ebpfio} moves code from userspace into the kernel, cutting on crossings between user- and kernel-space.
However, the eBPF path has its own overheads which increase with, e.g., more helper calls~\cite{miano23sketchinginebpf}.
Combined with eBPF lacking userspace optimization strategies like SIMD~\cite{miano23sketchinginebpf}, it creates a tradeoff: the gain by avoiding user-to-kernel-space transitions may not be worth the reduced compute capabilities.

Third, offloads typically use different hardware resources than the original, non-offload paths---meaning that any unused original resources may still be useful for the task.
This is exacerbated by offloads with single-digit speedups, as using the aggregated resources could yield significant benefits.
An example is full TCP offloads that degrade performance for connections that find the NIC memory full (\cref{sec:intro}).

\subparagraph{RDMA writes: latency degrades with cache misses}
We show\\that offload costs increase significantly for certain workloads by diving into RDMA writes, the use case of this paper.
RDMA writes allow an initiating host to transfer data to the memory of a remote target host, without requiring any involvement from the target's CPU, aside from e.g., memory registrations at setup time.
They are widely used, e.g., in key-value systems~\cite{kalia14herd}.

Consider the lifetime of an RDMA write, focusing on the target's end.
The target has already allocated in its DRAM memory regions (by default, backed by 4~KB pages in Linux) and has communicated the necessary metadata for the initiator to RDMA-write to them.
The initiator's CPU posts an RDMA write to a certain \emph{virtual} remote memory address within those regions.
The RNIC at the initiator's side (iRNIC) transmits the request, which eventually arrives at the target's RNIC (tRNIC).
To complete the write, the tRNIC must copy the data to the \emph{physical} address that corresponds to the virtual address.
This requires the tRNIC to have the virtual-to-physical address translation in its Memory Translation Table (MTT) cache.
So, upon an MTT cache miss, the tRNIC must wait for the translation to be fetched from memory over the PCIe, which interconnects NICs and CPU/DRAM.
This adds extra PCIe traversals, inflating the latency for the write to be visible in the target's memory.

We evaluate the cost of cache misses at the tRNIC, using the same experimental setup as in \cref{sec:eval}.
\Cref{fig:effectmrszipfian}, orange line, shows the results.
With 1 memory region (overall 4~KB), the average RTT latency is $\sim$2.6~$\mu$s, as there are no MTT capacity misses at the tRNIC.
Increasing the memory regions accessed by writes to $2^{20}$ (overall 4~GB) increases the average latency to $\sim$5.1~$\mu$s, i.e., a $\sim$2x increase.
This is because most writes now experience MTT capacity misses at the tRNIC and must additionally wait for the tRNIC to fetch the translations over PCIe before being written to memory.

\begin{tcolorbox}[myparagraphbox]
\paragraph{Idea 1: Unload part of an offloaded task}
We can accelerate offloaded tasks by (partially) ``unloading'' them: splitting the task into parts and moving back to the non-offload path only the parts that reduce the offload cost.
Intuitively, partially unloading a task aims to achieve the best of fully offloading and not offloading at all.
\end{tcolorbox}

For example, the ``unload path'' for RDMA writes would both avoid expensive TCP stack traversals and cut on expensive PCIe traversals.

\paragraph{Problem 2: Dynamic workloads lead to uncertain unload performance gains}
A workload consists of a mix of task characteristics, each potentially with a different offload cost.
Even worse, the offload cost depends on system state, e.g., what is in the cache, which changes over time.
Further, it might even make sense to pay the offload cost once to get over-compensated by future tasks, e.g., writes to heavy-hitter memory regions that reuse the cache.
At the same time, unloading does not come for free.
There can be two costs to it:
a constant one for keeping the semantics of the unload/offload paths the same, and a variable one that depends on the task characteristics, e.g., request size.
How do we know which tasks to unload and which ones to keep on the original, offload path to accelerate offloads?

\subparagraph{RDMA writes: blind unloading can worsen performance}
Consider two sequential RDMA writes:
Write 1 targets an unpopular memory region whose virtual-to-physical translation is not in the tRNIC cache.
Write 2, however, targets a heavy-hitter memory region whose translation is in the tRNIC cache.
There are four unloading options and the challenge is that some of them deliver worse performance than always offloading:
If we keep write 1 offloaded but unload write 2, on top of the offload cost to fetch the translation for write 1, we need to pay the unload cost for write 2---making this choice worse than keeping both writes offloaded.
The better choice is to unload write 1 (to avoid the extra PCIe traversals of the offload path) and keep write 2 offloaded (to benefit from caching on the offload path).

\begin{tcolorbox}[myparagraphbox]
\paragraph{Idea 2: Monitor tasks/system state and dynamically route tasks to offload/unload paths}
By rerouting at the granularity of tasks, we can use the most appropriate path based on the system state and task's resource (\emph{what}), location (\emph{where}), and processing (\emph{how}) requirements.
\end{tcolorbox}

Rerouting tasks across the unload/offload paths has two challenges:
First, as these decisions are on the critical path, which is already highly optimized, they must be fast and simple enough to avoid introducing overhead that offsets the unloading benefits.
Second, they need to be able to act despite the offload cost depending on hidden or remote state, which is challenging to retrieve accurately and timely.

\paragraph{Problem 3: Blurry application/offload boundaries hinder dynamic unloading}
If unloading was a static, a priori decision, we could rewrite applications to always use or not use the unload path for a certain task.
As argued, however, unloading decisions must happen dynamically at runtime to reap the gains.
Given the tight integration of offload paths into the application code, how can we ensure that this is possible without breaking the existing offload interface? How can we ensure that the results of dynamically unloaded tasks are properly communicated to the application?
This is particularly challenging for offloads like one-sided RDMA operations where 1)~consumers of the offloaded tasks are often application logic that has nothing to do with networking but just accesses memory, and 2)~there is no notification on the target node that a one-sided RDMA operation completed\footnote{with the exception of RDMA write with immediate}, even less a way to notify the application that it was unloaded.

\subparagraph{RDMA writes: unloading can break application logic}
Consid-%
er the typical RDMA write case: the initiator writes something into a memory location; the application learns about the memory location and the completion of the transfer through some unrelated application code (e.g., another RDMA operation, or a predetermined agreement on which memory address it should poll).
Assuming now that an unload path uses a two-sided send/recv operation instead, the original application would have no clue when the transfer has completed or where it can find the result.

\begin{tcolorbox}[myparagraphbox]
\paragraph{Idea 3: Unload through the offload interface}
To maintain compatibility with the existing offload code and allow dynamic switching between offload and unload, we can keep the interface the same and guarantee that any unload path preserves the end-to-end offload semantics (such as the final target location).
\end{tcolorbox}

\newcommand{\data}{unload module\xspace}
\newcommand{\control}{decision module\xspace}
\newcommand{\Data}{Unload Module\xspace}
\newcommand{\Control}{Decision Module\xspace}

\input{figures/arch}

\section{Unloading for Faster RDMA}
\label{sec:arch}

We propose \sys, a preliminary architecture for accelerating RDMA write offloads by dynamically unloading them.

At a high level, \sys intercepts existing RDMA write requests and dynamically decides whether to unload them (\cref{fig:arch}).
Unloaded requests still use the RDMA infrastructure, but instead of trying to push the whole operation to the RNICs, they execute some of the offloadable parts using the remote CPU.
As a result, operations that would be slower when offloaded or those that consume RNIC resources for little to no gain can make space for more offload-critical operations and leverage the CPU's close proximity to DRAM for, e.g., fast address translation.

The rest of this section describes \sys's two main components:
1)~the \emph{\control} (\cref{sec:arch:control}), that monitors the system state, intercepts requests, and decides which ones should be unloaded,
and 2)~the \emph{\data} (\cref{sec:arch:data}), that receives requests and unload decisions, based on which it puts and completes the requests on the unload/offload paths.
For simplicity, we describe \sys from the perspective of one node being the initiator of all writes, but the same architecture extends to the case where both nodes are initiators.

\subsection{\Data}
\label{sec:arch:data}
The unload module runs at both the initiator and the target node's CPU.
Given an RDMA write request and its corresponding decision (\cref{sec:arch:control}), the unload module is responsible for completing the request via the selected path.
The offload path remains unchanged, so we focus on the unload one.

At a high level, the unload path replaces an RDMA write to a potentially random memory region in the target's memory with a write into a specific (buffer) remote region.
Then, the remote CPU is responsible for transferring the write from that buffer to its intended (target) destination.
This buffer is reused for multiple small writes; thus, it is expected to be MTT-cache-resident, in contrast with the input target, which the policy (\cref{sec:arch:control}) determined it would likely be a miss.

\paragraph{Setup}
In addition to the typical RDMA-write initializations (e.g., queue-pair construction), 
in \sys the participating nodes exchange metadata regarding the temporary buffer location.
The temporary buffer is private to each queue pair.

\paragraph{Changes at the initiator}
The unload module at the initiator unloads a write by 1)~replacing it with an \emph{RDMA write with immediate (RDMA writeImm)} operation, 2)~rerouting the writeImm to the next slot in the target's temporary buffer, and 3)~updating the local metadata about the buffer usage (i.e, where to write the next unloaded operation, if any).

The writeImm payload contains the destination address from the original RDMA write, followed by the original write payload.
Further, the initiator has to communicate to the target node the stag and length of the write.
To avoid additional transfer overhead, the stag is passed as an immediate value in the RDMA writeImm, and the actual write length is computed through the received work completion.

The change from write to writeImm is a necessity and the main unload cost: RDMA does not guarantee the order in which RDMA-written bytes will become visible to the target's CPU, thus polling wouldn't be able to ensure that all data is transferred, and the unload layer does not have any other signals to determine that the transfer completed.

\paragraph{Changes at the target}
For unloaded writes, the target has two tasks:
First, to achieve functional parity in terms of transferred data and update location, the target's CPU: 1)~waits for the RDMA writeImm to complete,
2)~verifies that the combination of target address, size, stag, and access permission is valid, and 
3)~copies the data from the temporary buffer to its final target location.
Effectively, the MTT miss at the target's RNIC is replaced by a lookup for the security check, a memcpy, and a potential TLB miss.

Second, to guarantee security parity, the address, size, stag, and permission metadata for each memory region registration are stored in uMTT, a \sys local map, and removed during de-registration.
The security check of the previous paragraph is performed via a lookup into this map.

\subsection{\Control}
\label{sec:arch:control}

\sys's decision module runs at the initiator node CPU.
At runtime, the decision module intercepts RDMA write requests.
For each request, it uses as input the request characteristics together with an \emph{unload policy} and \sys-internal metadata to decide whether to unload it.
Then, it forwards the RDMA request and unload decision to the \data.

\paragraph{Unload policy}
\sys accepts multiple unload policies: each policy describes 1)~conditions for \sys to unload an RDMA request based on the expected state of the local and remote RNIC, and 2)~statistics for \sys to monitor.

We have two proof-of-concept policies:
The first one is a hint-based policy that assumes the application knows and marks the requests that should be offloaded in the RDMA post.
For example, an application that knows to which remote memory pages it frequently writes can mark only those for offloading---expecting that writes will benefit from the offload path only if they can reuse the RNIC cache on the offload path.
The second frequency-based policy follows the same philosophy but does not assume that the application can supply the heavy-hitter pages.
Instead, \sys tracks them using the \emph{monitor} (see below) and reroutes requests to the least frequently accessed pages to the unload path.

\paragraph{Monitor}
Depending on the unload policy, \sys needs to maintain statistics to decide the optimal path for each request.
The hint-based policy does not require any statistics, but the frequency-based one does.
Next, we sketch a way to do that fast, as policies have strict performance requirements: policies have to execute for each RDMA request and provide an answer faster than the expected savings, which range from a few hundred nanoseconds to a few microseconds.

\sys can track remote page frequencies by maintaining an array of counters, one per remote page, and unload only small writes whose estimated target pages appear less frequently than a configurable relative frequency threshold.
When a request arrives, \sys increments the counter corresponding to the remote page that the RDMA operation will access.
As RDMA operations only specify the target remote address range, the initiator makes the simplifying assumption that all remote virtual addresses are backed by 4~KB pages and updates the counters of the associated pages.
So, for small transfers (less than a page), deciding whether to unload a request requires updating one counter and comparing it with the threshold.
Good thresholds can be determined out of the critical path by looking at the frequency distribution.

\section{Preliminary Evaluation}
\label{sec:eval}

We experimentally show the tradeoff between offload/unload across workload characteristics, and that adaptively switching the offload/unload achieves the best of both worlds.

\paragraph{Experimental setup}
We use two servers, each with an NVIDIA ConnectX-5 Ex RNIC.
The two RNICs are directly connected.
For each experiment we report the average latency of 5 million sequential 16~B (inlined) RDMA writes from one server to the other. Each write selects a  4~KB memory region from a discrete Zipfian distribution with 0.5 skew.

The RDMA completion notification guarantees do not allow measuring latency directly~\cite{kashyap19rdmapersistence}, thus, similar to prior work~\cite{kalia14herd}, we measure the round-trip time:
The time from the initiator to post an RDMA write to seeing locally a 32~bit response  (always to the same buffer, so no MTT capacity misses at the initiator RNIC) written by the target.

To capture a range of workloads, we vary the total work area targeted by the RDMA writes.
Specifically, we vary the number of non-contiguous memory regions from 1 to $2^{20}$.
All entries fit in the MTT cache for low region counts, while capacity misses increase with higher number of regions.

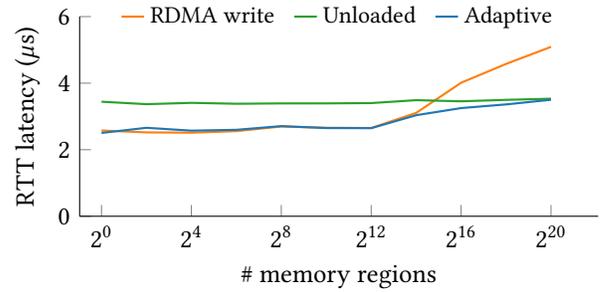
\begin{figure}[!t]
    \begin{tikzpicture}
    \begin{axis}[
        xmode=log,
        log basis x=2,
        xlabel={\# memory regions},
        ylabel={RTT latency ($\mu$s)},
        table/x=mrs,
        xmin=0.51,
    ]
    \addplot+ [label={RDMA write}] table [x=mrs, y={rdma_write}]{data/latency_vs_mrs_zipfian.txt};
    \addplot+ [label={Unloaded}] table [x=mrs, y={unloaded_rdma_write}]{data/latency_vs_mrs_zipfian.txt};
    \addplot+ [label={Adaptive}] table [x=mrs, y={adaptive}]{data/latency_vs_mrs_zipfian.txt};
    \end{axis}
    \end{tikzpicture}
    \caption{Unloading improves RTT latency by up to 31\%, and adaptive achieves the overall best RTT latency.}
    \label{fig:effectmrszipfian}
\end{figure}

\paragraph{To offload or to unload}
\Cref{fig:effectmrszipfian} shows the RTT latency when using typical, offloaded RDMA writes (orange line) versus unloading all of them with \sys (green line).
We make two observations:
First, as we increase the accessed memory regions from 1 to $2^{20}$ (i.e., from 4~KB to 4~GB), the unload path's latency stays almost unaffected ($\sim$3.4~$\mu$s)---showing that \emph{unloading indeed avoids MTT cache miss costs that typically appear for many memory regions}.
Second, while for few memory regions the offload path can reuse the cache and thus is faster than the unload path ($\sim$2.6~$\mu$s vs. $\sim$3.4~$\mu$s), \emph{when the working set no longer fits in the cache, the unload path is faster}.
E.g., for $2^{20}$ memory regions, the unload path's RTT latency is $\sim$3.5~$\mu$s compared to $\sim$5.1~$\mu$s for offloaded RDMA writes, i.e., a $\sim$31\% improvement.

\paragraph{Better to adaptively unload}
\Cref{fig:effectmrszipfian}, blue line, shows the potential of adaptively switching between the unload/offload path using the hint-based unload policy (that offloads only the top-4096 heavy-hitter memory regions).
Adaptively unloading matches the best latency across configurations, or has even lower latency than always taking either path when adaptive can take advantage of the RNIC cache.

\section{Discussion \& Future Directions}
\label{sec:future}

\paragraphnodot{Isn't RDMA all about networking without the remote CPU?}
While one of the advantages of RDMA's one-sided operations is bypassing the remote CPU for data-intensive transfers, RDMA applications often \emph{do} use the remote CPU for their application logic (e.g., for receiving completion notifications for a batch of RDMA operations and processing them).
With unloaded RDMA writes, part of the work is accelerated by pushing it back to the CPU that runs the rest of the application logic, while data-intensive transfers still use the RNIC to bound the CPU overheads.
Furthermore, unloaded RDMA writes still avoid TCP/IP stack traversals, one of the other core motivations of RDMA offloading.

\paragraphnodot{How do we ensure full compatibility between RDMA offload/unload interfaces?}
\sys already aligns with existing application expectations in terms of transferred data, update location, and security (\cref{sec:arch:data}).
However, it does not guarantee parity in terms of event completions/notifications and RDMA operation ordering (e.g., flushes).
While we believe this would be possible with our current software-only approach (e.g., event parity by using a secondary queue pair for unload events and generating the offload-expected notifications in the original completion queue), ordering parity would likely incur a performance penalty.

In contrast, pushing the unload module into the remote RNIC would unlock \sys's full benefits.
Specifically, in such an implementation, \sys's buffer would be provided to the target RNIC as part of the queue pair metadata.
Then, whenever the RNIC receives an RDMA write that \sys decided to unload,
the RNIC 1)~writes the RDMA write and its relevant metadata into the buffer, and 2)~generates the notification that \sys currently generates through the write with immediate, for the CPU to 3)~complete the unload operation by doing the final copy.
In addition, such an implementation would alleviate the need for the initiator to estimate the remote RNIC cache content---at the expense of requiring hardware changes that are known to have a slow adoption rate~\cite{alcoz20sppifo,barkalov19foundationsembedded}.
Lastly, with such a hardware implementation, \sys would be able to generate the same notification and ordering requirements as the original operation, as it would have access to internal RNIC mechanisms. 

\paragraphnodot{How do we generalize to other offloads?}
We have focused on how unloading accelerates RDMA writes.
However, \sys's motivation and architecture also apply to other offloads.
For example, consider an application that offloads packet processing from userspace to the eBPF virtual machine (VM).
The principles of \cref{sec:motivation} extend to this offload: the execution of the eBPF program in kernelspace means that its location changed to be closer to kernel structures, instead of application structures, and the eBPF VM changes the execution profile, e.g., because its instruction set does not have SIMD.
This introduces overheads, e.g., when the offload program interacts with the userspace frequently, or processing some incoming packets becomes demanding.
Reapplying \sys's core ideas would allow detecting cases where this overhead surpasses the benefits, and dynamically unloading the processing of some packets back to userspace.

Finally, the broad applicability of the presented unload techniques calls for an automated method to unload offloaded applications.
To that end, we plan to 1)~investigate approaches to automatically generate efficient unload paths using code synthesis and generative techniques, and 2)~combine them with a toolset for low-overhead monitoring that will drive the unload decisions.

\section{Related Work}
\label{sec:related}

This section reviews related work on RDMA, similar concepts (DMA), and other offloads (eBPF).

\paragraph{Accelerating RDMA}
Huge pages require fewer MTT entries to map the memory---reducing MTT misses~\cite{guo16rdmaatscale} and PCIe traversals but also leading to memory wasting/bloating~\cite{novakovic19storm,park20perforatedpage}.
Storm~\cite{novakovic19storm} uses physical segments in combination with contiguous memory allocators to reduce the MTT entries---avoiding the need for translations but requiring applications to register segments with the OS.
Our work offers a new point in this space---accelerating RDMA by dynamically deciding when to unload it.

\paragraph{Finding problematic RDMA workloads}
Collie~\cite{kong22collie} finds RDMA workloads that trigger lower-than-expected throughput or redundant backpressure on RDMA traffic.
We envision extending \sys with approaches similar to Collie's space exploration to automatically find workloads for unloading.

\paragraph{Accelerating DMA}
Prior work accelerates DMA for TCP traffic:
It addresses security/synchronization overheads upon unmapping DMA memory by redirecting DMAs to special buffers that are never unmapped~\cite{markuze16trueiommuprotection}, or reduces translation overheads by redirecting DMAs to huge pages already present in the IOTLB~\cite{farshin23overcomingiotlbwall}.
Unlike TCP, where the receiver controls the target buffers, we focus on accelerating RDMA traffic, where the application controls the target buffers.

\paragraph{Automatic eBPF offloads}
Shahinfar et al.~\cite{shahinfar23automaticoffloadebpf} propose a compiler design that automatically splits userspace programs to a userspace and an offloaded eBPF part, based on both feasibility and performance benefits.
We take the opposite approach by starting from the offload.
This creates an opportunity: there are fewer feasibility concerns as the two ends of the spectrum (no or full offload) are possible/implemented, so we can focus on performance-related challenges.

\section{Conclusion}

We proposed accelerating offloads by making them bidirectional: amenable to dynamically switch between the offload ``forward path''---that moves a task from (closer to) the application layer to (closer to) the network/link layer---and the unload ``backward path''---that moves (part of) the task from the network/link back to the application layer.
We showed the potential of bidirectional offloads by taking a first step to applying them to RDMA writes: we showed how to construct the unload path, how to stay compatible with the offload path in terms of transferred data, write location and security, and how to decide which path each RDMA write should take to improve the offload’s performance.

\paragraph{Ethical issues} This paper does not raise any ethical issues.



\balance
\bibliographystyle{ACM-Reference-Format} 
\bibliography{hotnets25-template}

\end{document}